\documentstyle[multicol,eqsecnum,prd,aps,graphics]{revtex}    
\begin{document}
\draft
\renewcommand{\baselinestretch}{1.2}
\def\dis{\displaystyle}
\newcommand{\beq}{\begin{equation}}
\newcommand{\eeq}{\end{equation}}
\newcommand{\beqa}{\begin{eqnarray}}
\newcommand{\eeqa}{\end{eqnarray}}
 
\def\dis{\displaystyle}
 
\def\nue{{\nu_e}}
\def\anue{{\bar{\nu_e}}}
\def\numu{{\nu_{\mu}}}
\def\anumu{{\bar{\nu_{\mu}}}}
\def\nutau{{\nu_{\tau}}}
 
\def\anutau{{\bar{\nu_{\tau}}}}
\def\l{\left}
\def\r{\right}
\def\lan{\langle}
\def\ran{\rangle}
\newcommand{\dm}{\mbox{$\Delta{m}^{2}$~}}
\newcommand{\st}{\mbox{$\sin^{2}\theta$~}}      
 
 
 
\title{ 
Detection Rates for Kaluza-Klein Dark Matter}
\vskip 20pt

\author{Debasish Majumdar\thanks{ 
E-mail address: debasish@theory.saha.ernet.in}}
\address{Theory Division, Saha Institute of Nuclear Physics,\\  
1/AF Bidhannagar, Kolkata 700064, India. }
\date{\today}
\maketitle 
\begin{abstract}
We consider the lightest Kaluza-Klein particle at N=1 mode (LKP) of 
universal extra
dimension to be the candidate for Dark Matter and predict the detection
rates for such particles for Germenium and NaI detectors. We have also 
calculated the nature of annual modulation for the signals in these two 
types of detectors for LKP Dark Matter. The rates
with different values of speed of solar system in the Galactic rest frame 
are also evaluated.
\vspace{1pc}
\end{abstract}
\maketitle
\begin{multicols}{2}
\narrowtext
\section{Introduction}
There are strong indirect evidence (gravitational) from various observations
like velocity curves of spiral galaxies, gravitational lensing etc., in 
favour of the existence of enormous amount of invisible, nonluminous matter
or Dark Matter in the universe. This Dark Matter constitutes more than 90\%
of the matter in the universe. Although the constituents of Dark Matter 
still remain a mystery, the indirect evidence suggests that large part of 
Dark Matter should be of non-Baryonic in nature. They are stable, heavy, 
non relativistic (Cold Dark Matter or CDM) and are weakly 
interacting. Therefore they are often known as Weakly Interacting Massive 
Particles or WIMPs. 

Particle physics has suggested a number of options for the candidates of 
these non baryonic Cold Dark Matter. Currently the most popular candidate 
is a lightest supersymmetric particle (LSP) neutralino ($\chi$) which of 
course is not a Standard Model particle. In Minimal Supersymmetric Standard 
Model or MSSM, the LSP is a neutralino ($\chi$). Neutralino is a Majorana 
fermion and it is a superposition of supersymmetric partners of neutral 
U(1) and SU(2) gauge bosons and neutral Higgs bosons. The conservation 
of R-parity ensures that the LSP is a stable particle. In literature 
there are a lot of work where neutralino in MSSM and supergravity models
is consudered as a dark matter candidate 
\cite{bottino1,bottino3,howard,edsejo,utpal}. In a recent work 
neutralino LSP as a DarK Matter candidate from minimal Anomaly Mediated 
Supersymmetric Breaking (mAMSB) model has been addressed \cite{dm}.  

In the present work we consider a bosonic particle, unlike the fermionic 
SUSY particle to be the candidate for Dark Matter. This particle is a 
lightest Kaluza-Klein (KK) particle or LKP in universal extra dimension (UED)
\cite {cheng1,cheng2}. The Kaluza-Klein theory \cite {kk} 
which is a 5-dimensional theory with 4 space-time dimensions and 
1 extra space dimension, the extra dimension
(5$^{th}$ dimension) is assumed to be compact. The fifth dimension has 
the topology of a circle with the compactification radius  
of the order of Planck scale. The topology of this 5-Dimensional space-time
is $R^4 \times S^1$ and the fifth coordinate $y$ is periodic with 
$0 \le my \le 2\pi$, $m$ being inverse of the radius of the circle.
This periodicity of the extra dimension enables one to obtain an infinite tower
of fields by making Fourier expansion in the fifth coordinate. This tower 
is known as Kaluza-Klein tower. The Universal Extra Dimension model
is one where all Standard Model particles can propagate in extra dimensions
of size $R \sim {\rm TeV}^{-1}$. In UED therefore every standard model particle
has a KK partner in KK tower. In the present case UED scenario is considered
with only one extra dimension. The parity conservation of KK particles 
ensures that the lightest KK particle or LKP is stable and thus can be a 
candidate for cold Dark Matter. 

The direct detection of WIMP Dark Matter by a terrestrial detector
uses the principle of elastic scattering off detector nuclei. But this 
is a difficult task as the WIMP-nucleus interaction is very feeble. 
The energy deposited by a WIMP of mass few GeV to 1 TeV on a detector 
nucleus is not more than 100 keV. Hence for direct detection of WIMP 
the detector has to be of low energy threshold and of low background. 

Presently, there are several experiments engaged in direct detection 
of Cold Dark Matter. The direct WIMP search by 
DAMA/NaI collaboration \cite{dama}  
with $\sim 100$ kg NaI(TI) set up at Gran Sasso in Italy looks for 
{\it annual modulation signature} of WIMP. Due to the earth motion 
around the sun, the earth bound detector will experience a larger WIMP
flux in the month of June when the earth's rotational the velocity 
adds up to the velocity of solar system in Galaxy and minimum in December 
when these two velocities are antiparallel. 
The DAMA collaboration claimed to have detected this 
annual modulation of WIMP through their direct WIMP detection experiments.
Their analysis suggests possible presence of Dark Matter with mass 
around 50 GeV.
The Cryogenic Dark Matter Search or CDMS detector employs low temperature 
Ge and Si as detector materials to detect WIMP's via their elastic
scattering off these nuclei \cite{cdms}. This is housed in a 10.6 m 
tunnel ($\sim 16$ m.w.e) at Stanford Underground Facility 
beneath the University of Stanford.  
Although their direct search results are compatible with some regions 
of 3-$\sigma$
allowed regions for DAMA analysis, it excludes DAMA results if standard 
WIMP interaction and a standard Dark Matter halo is assumed. The 
EDELWEISS Dark Matter search experiment which also uses cryogenic
Ge detector at Frejus tunnel, 4800 m.w.e under French-Italian Alps 
observed no nuclear recoils in the fiducial volume \cite{edelweiss}. 
The lower bound of recoil energy in this experiment was 20 keV. 
The Heidelberg Dark Matter Search (HDMS) uses in their inner detector,
highly pure $^{73}$Ge crystals \cite{klapdor0} and with a very 
low energy threshold. They have recently made available their
26.5 kg day analysis. The recent low threshold experiment GENIUS
(GErmenium in liquid NItrogen Underground Setup) \cite{klapdor1} 
at Gran Sasso tunnel in Italy
has started its operation. Although a project for $\beta\beta$-decay search, 
due to its very low threshold (and expected to be reduced futher) 
GENIUS is a potential detector for WIMP direct detection 
experiments and for detection 
of low energy solar neutrinos like pp-neutrinos or $^7$Be neutrinos. 
In GENIUS experiment highly pure $^{76}$Ge is used as detector material.
For Dark Matter search 100 kg. of the detector material is suspended 
in a tank of liquid nitrogen. In Fig. 3 of Ref. \cite{klapdor1} the 
projected limit for WIMP mass $m_\chi$ and scalar cross-section for this
detector is given. In the present work we have calculated the prediction 
for rates for the case two types of detector material namely  
$^{76}$Ge and NaI.
    
The theoretical predictions of rates has been addressed by several 
authors \cite{bottino2,belli,brhlik}. Although in Ref. \cite{bottino2}
rates for detectors with various detecting material is given.  
In Refs. \cite{belli,brhlik}, the emphasis was mainly to analyse 
DAMA/NaI data. 

The paper is organised as follows. In section 2 we give the theory for 
calculation of detection rates for Ge and NaI detectors. 
The actual calculation of differential rates for 
$^{76}$Ge detector and NaI detector for various choices of WIMP mass and 
other parameters are discussed in section 3. In Section 4 we give some
concluding remarks.

\section{Theory} 

Differential detection rate of WIMPs per unit detector mass can be 
written as 
\begin{equation}
\frac {dR} {d|{\bf q}|^2} = N_T \Phi \frac {d\sigma} {d|{\bf q}|^2} \int f(v) dv
\end{equation}

where $N_T$ denotes the number of target nuclei per unit mass of the detector,
$\Phi$ being the WIMP flux and $v$ is WIMP velocity in the reference 
frame of earth and $f(v)$ is the distribution of this velocity. The integration
is over all possible kinematic configurations in the scattering process.
In the above, $|\bf q|$ is the momentum transferred to the nucleus in 
WIMP-nucleus scattering. Nuclear recoil energy $E_R$ is

\beqa
E_R &=& |{\bf q}|^2/2m_{\rm nuc} \nonumber \\
    &=& m^2_{\rm red} v^2 (1 - \cos\theta)/m_{\rm nuc}  \\
m_{\rm red} &=& \frac {m_\chi m_{\rm nuc}} {m_\chi + m_{\rm nuc}}  
\eeqa
where $\theta$ is the scattering angle in WIMP-nucleus centre of momentum
frame, $m_{\rm nuc}$ is the nuclear mass and $m_\chi$ is the WIMP mass. 

Now expressing $\Phi$ in terms of local WIMP density $\rho_\chi$,  WIMP
velocity $v$ and WIMP mass $m_\chi$ and writing $|{\bf q}|^2$ in terms
of nuclear recoil energy $E_R$ with noting that $N_T = 1/m_{\rm nuc}$, 
Eq. (1) takes the form 

\beq
\frac {dR} {dE_R} = \frac {\rho_\chi} {m_\chi} 2 \frac {d\sigma} 
{d |{\bf q}|^2} \int_{v_{min}}^\infty v f(v) dv
\eeq

The WIMP-nucleus (or WIMP-nucleon) scattering cross-section has two parts,
namely spin-independent or scalar cross-section and spin dependent 
cross-section. Here we make the assumption that scalar cross section 
dominates over the spin dependent cross-section \footnote {For bosonic 
KK dark matter scalar cross-section effects dominate \cite{cheng1}}.
Following Ref. \cite{jungman} the WIMP-nucleus differential cross-section 
for the scalar interaction can be written as
\beq
\frac {d\sigma} {d |{\bf q}|^2} = \frac {\sigma_{\rm scalar}} 
{4 m_{\rm red}^2 v^2} F^2 (E_R) \,\,\, .
\eeq
In the above $\sigma_{\rm scalar}$ is WIMP-nucleus scalar cross-section
and $F(E_R)$ is nuclear form factor given by \cite{helm,engel}
\beqa
F(E_R) &=& \l [ \frac {3 j_1(qR_1)} {q R_1} \r ] {\rm exp} \l ( \frac {q^2s^2}
{2} \r ) \\
R_1 &=& (r^2 - 5s^2)^{1/2} \nonumber \\
r &=& 1.2 A^{1/3} \nonumber
\eeqa 
where thickness parameter of the nuclear surface is given by $s \simeq 1$ fm,
$A$ is the mass number of the nucleus and $j_1(qR_1)$ is the spherical 
Bessel function of index 1. 

For the distribution $f(v_{\rm gal}$ of WIMP velocity $v_{\rm gal}$ 
with respect to 
Galactic rest frame, a Maxwellian form is considered here. The 
velocity $v$ (and $f(v)$) with respect to earth rest frame can then be obtained
by making the transformation
\beq
{\bf v} = {\bf v}_{\rm gal} - {\bf v}_\oplus
\eeq
where $v_\oplus$ is the velocity of earth with respect to Galactic rest 
frame and is given by 
\beqa
v_\oplus &=& v_\odot + v_{\rm orb} \cos\gamma \cos \l (\frac {2\pi (t - t_0)}
{T} \r ) 
\eeqa
In Eq. (2.8), $T = 1$ year time period of earth motion around the sun,
$t_0 = 2^{\rm nd}$ June, $v_{\rm orb}$ is earth orbital speed and 
$\gamma \simeq 60^o$ is the angle subtended by earth orbital 
plane at Galactic plane. The speed of solar system $v_\odot$ in the 
Galactic rest frame is given by,
\beqa
v_\odot &=& v_0 + v_{\rm pec}
\eeqa
where $v_0$ is the circular velocity of the Local System at the position of
Solar System and $v_{\rm pec}$ is speed of Solar System with respect to 
the Local System. The latter is also called peculiar velocity and its value
is 12 km/sec. Although the physical range of $v_0$ is given by 
\cite{leonard,kochanek}
$170\,\, {\rm km/sec} \leq v_0 \leq 270$ km/sec (90 \% C.L.), in the present
work we consider the central value of $v_0 = 220$ km/sec. Here we mention 
that Eq. (2.8) is the origin of annual modulation of WIMP signal reported 
by DAMA/NaI experiment \cite{dama}.

Defining a dimensionless quantity $T(E_R)$ as, 
\beq
T(E_R) = \frac {\sqrt {\pi}} {2} v_0 \int_{v_{\rm min}}^\infty \frac {f(v)}
{v} dv\,\,
\eeq
and noting that $T(E_R)$ can be expressed as \cite{jungman}

\beq
T(E_R) = \frac {\sqrt {\pi}} {4v_\oplus} v_0 \l [ {\rm erf} \l ( \frac 
{v_{\rm min} + v_\oplus} {v_0} \r ) -  {\rm erf} \l ( \frac
{v_{\rm min} - v_\oplus} {v_0} \r ) \r ]
\eeq
we obtain from Eqs. (2.4) and (2.5)
\beqa
\frac {dR} {dE_R} &=& \frac {\sigma_{\rm scalar}\xi\rho_0} {4v_\oplus m_\chi 
m_{\rm red}^2} F^2 (E_R) \l [ {\rm erf} \l ( \frac 
{v_{\rm min} + v_\oplus} {v_0} \r ) \r. \nonumber \\ 
&&\l. - {\rm erf} \l ( \frac
{v_{\rm min} - v_\oplus} {v_0} \r ) \r ]
\eeqa
In the above, $\rho_0$, is the total local Dark Matter 
density geneally taken to be 0.3 GeV/cm$^3$
and $\xi = \rho_\chi / \rho_0$. The above expression for differential rate 
is for a monoatomic detector like Ge but it can be easily extended for 
a diatomic detector like NaI as we will see later.  

In the present case, the lightest KK state or LKP, $B^1$, in simplest UED is 
considered to be the candidate for WIMP Dark Matter \cite{cheng1}. The spin 
independent cross-section for scattering of $B^1$ with mass $m_{B^1}$ 
off nucleon or nucleus is given by \cite{cheng1}
\beqa
\sigma_{\rm scalar} &=& \frac {m_N^2} {4\pi (m_{B^1} + m_N)^2} 
\l [ Zf_p + (A - Z)f_n \r ]^2 \\
f_p & = & \dis\sum_{\rm all\,\, q} 
\frac {\beta_q + \gamma_q} {m_q} m_p f^p_{T_q} \nonumber \\
f_n & = & \dis\sum_{\rm all\,\, q} 
\frac {\beta_q + \gamma_q} {m_q} m_n f^n_{T_q} \nonumber \\
\beta_q & = & m_q \frac {e^2} {\cos^2\theta_W} 
\l [ Y^2_{qL} \frac {m_{B^1}^2 + m_{q_L^1}^2} { (m_{q_L^1}^2 - m_{B^1}^2)^2 }
+ ({\rm L} \rightarrow {\rm R}) \r ] \nonumber   \\
\gamma_q & = & m_q \frac {e^2} {2 \cos^2\theta_W} \frac {1} {m_h^2} \nonumber
\eeqa 
In Eq. (2.13), $m_N$ is the nuclear or nucleon mass, $q$ denotes the quarks, 
$Y$ is the hypercharge and $m_h$ is Higgs mass. For the mass 
$m_{q^1}$ we follow 
Ref. \cite{cheng1} and define a degeneracy parameter 
$d = (m_{q^1} - m_{B^1})/m_{B^1}$
and then assign different values of $d$ for the actual calculation of 
WIMP detection rates.
Needless to mention here that in the present calculation 
$m_\chi \equiv m_{B^1}$.

The measured response of the detector by the scattering of WIMP off detector
nucleus is in fact a fraction of the actual recoil energy. Thus, the actual 
recoil energy $E_R$ is quenched by a factor $qn_X$ (different for different 
nucleus $X$) and we should express differential rate in Eq. (2.12) in terms of 
$E = qn_XE_R$. For Ge detector the expected energy spectrum per energy bin 
can be expressed as 
\beq
\frac {\Delta R} {\Delta E} (E) = 
\dis\int^{(E + \Delta E)/qn_{\rm Ge}}_{E/qn_{\rm Ge}}
\frac {dR_{\rm Ge}} {dE_R} (E_R) \frac {dE_R} {\Delta E}
\eeq
and for a diatomic detector NaI, the above expression takes the form
\beqa
\frac {\Delta R} {\Delta E} (E) &=& 
a_{\rm Na} \dis\int^{(E + \Delta E)/qn_{\rm Na}}_{E/qn_{\rm Na}}
\frac {dR_{\rm Na}} {dE_R} (E_R) \frac {dE_R} {\Delta E}  \nonumber \\
&+&a_{\rm I} \dis\int^{(E + \Delta E)/qn_{\rm I}}_{E/qn_{\rm I}}
\frac {dR_{\rm I}} {dE_R} (E_R) \frac {dE_R} {\Delta E} 
\eeqa
where $a_{\rm Na}$ and  $a_{\rm I}$ are the mass fractions of Na and I 
respectively in a NaI detector and are given by (see Table 2)
$$
a_{\rm Na} = \frac {m_{\rm Na}} 
{m_{\rm Na} + m_{\rm I}} = 0.153 \,\,\,\,\, 
a_{\rm I} = \frac {m_{\rm I}} 
{m_{\rm Na} + m_{\rm I}} = 0.847  
$$

In this work theoretical prediction of the differential rate is calculated 
with $\Delta E = 1$ keV.

\section{Calculation of WIMP detection Rates for Germenium and NaI detectors}
In order to calculate the theoretical predictions for the variation of detected 
signals with $E$, we use Eqs. (2.12 - 2.15) alongwith Eqs. (2.3, 2.6-2.9). 
The scalar cross-section 
for the direct detection of LKP Dark Matter is estimated using 
Eq. (2.13). For $f^x_{T_q}$ ($x = p$ or $n$), we adopt 
the central values of these quantities given in \cite{cheng1}. 
Thus, $f^p_{T_u} = 0.020$,
$f^p_{T_d} = 0.026$, $f^p_{T_s} = 0.118$, $f^n_{T_u} = 0.014$,
$f^n_{T_d} = 0.036$, $f^n_{T_s} = 0.118$ and $f^{x}_{T_{c,b,t}} =
2(1 - f^{x}_{T_u} -  f^{x}_{T_d} -  f^{x}_{T_s})/27$. We have 
adopted four different values for degeneracy parameter 
$d (= (m_{q^1} - m_{B^1})/m_{B^1})$ and 
they are $d = 0.05$, 0.2, 0.3, and 0.4.
\vskip 2mm
For the actual calculations, 
we need to know among other things the value of $\xi$, the fraction of 
local WIMP density $\rho_\chi$ with respect to the total local Dark 
Matter density $\rho_0$. This is usually 
estimated by calculating the relic WIMP density, $\Omega_\chi h^2$ 
where $\Omega_\chi = \rho_\chi/\rho_c$, ($\rho_c$ is the critical density 
of the universe) and $h$ is related to the value of Hubble parameter at 
the present epoch.  This in 
turn depends on WIMP-WIMP annihilation cross-section (e.g. in
\cite{bottino1,edsejo}). In this first calculation of WIMP detection rates 
with a bosonic Kaluza-Klein particle considered as a candidate for WIMP,
we adopt the following procedure for the estimation of $\xi$. DAMA/NaI 
direct Dark Matter detection experiment has furnished their results of 
annual modulation signal signature of WIMP in the form of allowed contours
in the parameter space of $m_\chi$ and $\xi \sigma_{\rm scalar}^p$ 
\cite{dama}. Recently started GENIUS experiment (uses pure $^{76}$Ge as 
detector material) \cite{klapdor1} has predicted the variation of 
$\sigma_{\rm scalar}^p$ for different values of WIMP mass $m_\chi$. 
Here $\sigma_{\rm scalar}^p$ is WIMP scalar cross section on proton
normalised to square of the nucleon number ($A^2$). 
Now, $\xi$ is estimated by first calculating  $\sigma_{\rm scalar}^p$
from Eq. (2.10) for different $m_\chi$'s (for different fixed values of $d$).
From the values of $\xi \sigma_{\rm scalar}^p$ predicted by the 
experiments (as the ones discussed above) for the same $m_\chi$'s,  
one can estimate the value of $\xi$ for different $m_\chi$ (and also for 
different $d$) using the ratio $\xi = \frac 
{(\xi \sigma_{\rm scalar}^p)_{\rm Expt.}} {\sigma_{\rm scalar}^p}$.
In order to calculate $\sigma_{\rm scalar}^p$ for KK 
Dark Matter from Eq. (2.10),
we make the approximation $f_p = f_n$.
Now replacing $m_N$ by $m_p$, the mass of the proton, in Eq. (2.10), 
and dividing the expression by $A^2$ gives $\sigma_{\rm scalar}^p$ 
for LKP Dark Matter. We have calculated the variation of  
$\sigma_{\rm scalar}^p$ with $m_\chi$ for KK Dark Matter (LKP) 
for four different
values of $d$ ($d = 0.05$, 0.1, 0.2, 0.4). The results are plotted in Fig. 1.
Also plotted in Fig. 1 are the 3$\sigma$ C.L. DAMA/NaI allowed region in 
$\xi \sigma_{\rm scalar}^p$ - $m_\chi$ plane for $v_0 = 220$ km/sec 
(Fig. 4b in Ref. \cite{dama}) and the estimated limit of the variation 
$\sigma_{\rm scalar}^p$ with $m_\chi$ from GENIUS detector with 
100 kg. of natural $^{76}$Ge (Fig. 3 of Ref. \cite{klapdor1}). Here we 
adopt the following prescription for estimation of $\xi$. If the ratio 
$\xi < 1$ then we take the actual calculated value of $\xi$ from 
Fig. 1. Otherwise if $\xi \geq 1$ then we simply take $\xi = 1$. With 
this prescription, we see from Fig. 1, that if we use DAMA/NaI plot
for the estimation of $\xi$ then for most of the cases $\xi = 1$. As this 
is an unlikely scenario, we consider instead the GENIUS (100 kg natural Ge)
plot and estimate $\xi$ (using the ratio mentioned above) for different 
values of $m_\chi$ with different fixed values of $d$. The values of $d$
are chosen from the range given in Ref. \cite{cheng1}. We have taken 
four different values of degeneacy parameter $d$ ($d = 0.05$, 0.1, 0.2, 0.4)
and for each value of $d$ we have taken five values of $m_\chi$ ($m_chi =
30$ GeV, 50 GeV, 80 GeV, 100 GeV, 150 GeV) for estimation of the ratio $\xi$.
The results are furnished in a tabulated form in Table 1.

\begin{center}
\begin{tabular}{|c|c|c||c|c|c|}
\hline
$d$ & $m_{\chi}$ & $\xi$ & $d$ & $m_{\chi}$ & $\xi$ \\
     & (GeV)      &      &      & (GeV)      &      \\
\hline
     & 30.0       & 2.16e-05 &  & 30.0 & 3.43e-04 \\
     & 50.0       & 2.26e-04 &  & 50.0 & 3.56e-03 \\
0.05     & 80.0       & 4.17e-03 & 0.1 & 80.0 & 6.49e-02 \\
     & 100.0      & 1.58e-02 &     & 100.0 & 2.42e-01 \\
     & 150.0      & 2.21e-01 &     & 150.0 & 1.0      \\
\hline
     & 30.0       & 5.35e-03 &  & 30.0 & 7.82e-02  \\
     & 50.0       & 5.43e-02 &  & 50.0 & 7.29e-01 \\
0.2     & 80.0       & 9.36e-01 & 0.4 & 80.0 & 1.0 \\
     & 100.0      & 1.0      &  & 100.0 & 1.0 \\
     & 150.0      & 1.0      &  & 150.0 & 1.0   \\
\hline
\end{tabular}
\end{center}

\begin{description}
\item{\small \sf Table 1:} {\small \sf Estimated values of $\xi$, the 
ratio of local WIMP density and total local Dark Matter density for 
different WIMP mass $m_\chi$ and degeneracy parameter $d$ (see text).}
\end{description}
 
\noindent From Table 1 we see that $\xi$ increases with increase of the value of
$d$ as also the mass $m_\chi$. This feature is reflective of 
the fact that $\sigma^p_{\rm scalar}$ decreases with degeneracy $d$ and 
$m_\chi$ for the range of values chosen for present study. 

Having obtained $\xi$ from Table 1, we now proceed to calculate 
the WIMP detection rates for a Germenium detector that uses $^{76}$Ge as 
detector material. These rates are computed using Eqs. (2.4 - 2.14)
in case of four degeneracy parameters mentioned earlier
($d = 0.05$, 0.1, 0.2, 0.4) and for five values of 
$m_\chi$'s 30 GeV, 50 GeV, 80 GeV, 100 GeV 
and 100 GeV for each value of $d$ (see Table 1). 
The WIMP - nucleus scalar cross sections are evaluated by replacing
$m_N$ in Eq. (2.13) by 
mass of the nucleus $m_{\rm nuc}$ and use the values of $f_p$ and $f_n$
as given above.    
The nuclear mass $m_{\rm nuc}$ is calculated using the relation 
$m_{\rm nuc} = (Zm_p + (A - Z)m_n) + \Delta$, where $m_p$ and $m_n$ are
respectively the mass of the proton and neutron. The mass excess 
is denoted as $\Delta$ (electron masses are neglected). Table 2 
gives the values of $\Delta$ (in MeV)
and calculated values of $m_{\rm nuc}$ for $^{76}_{32}{\rm Ge}$,
$^{23}_{11}{\rm Na}$, $^{127}_{53}{\rm I}$ nuclei.  

\begin{center}
\begin{tabular}{|c|c|c|c|}
\hline
Nucleus & $\Delta$ & Ref. & $m_{\rm nuc}$\\
        &  (MeV)   &      & (GeV)  \\
\hline
$^{76}_{32}{\rm Ge}$ & -73.2127 & \cite{isotope} & 71.2921 \\
&&& \\
$^{23}_{11}{\rm Na}$ & -9.5295 & \cite{isotope} & 21.5862 \\
&&& \\
$^{127}_{53}{\rm I}$ & -88.988 & \cite{isotope}  & 119.1668 \\
\hline
\end{tabular}
\end{center}

\begin{description}
\item{\small \sf Table 2:} {\small \sf Mass excess $\Delta$ (in MeV) 
and calculated nuclear mass $m_{\rm nuc}$ for three nuclei used for WIMP 
detection}  
\end{description}

We first calculate differential rates $\frac {dR} {dE_R}$ (Eq. 2.12) for 
a wide range of values of recoil energy $E_R$ with $v_0 = 220$ km/sec.
The value of $t$ (day number) in Eq. (2.8) is taken to be $t_0$, i.e. 
the calculations are for 2$^{\rm nd}$ June. Then we use Eq. (2.14) 
for predicting  the observable rate $\frac {\Delta R} {\Delta E}$
(Eq. 14) in the units of kg$^{-1}$ day$^{-1}$ keV$^{-1}$.    
The quenching factor
for Ge, $qn_{^{76}{\rm Ge}} = 0.25$ \cite{bottino2} and $\Delta E = 1$ keV.

The results are shown in Fig. 2a to Fig. 2d. Each of the graphs (a - d)
in Fig. 2 is for rates for five different values of $m_\chi$ with a fixed
value of degeneracy $d$ (Table 1). For the sake of clarity of the plots, 
the upper limit of $\frac {\Delta R} {\Delta E}$ in all the four graphs
in Fig. 2 is truncated at 0.2  kg$^{-1}$ day$^{-1}$ keV$^{-1}$. Therefore in 
Fig. 3 we plot the same (rate vs $E$) for the case of $m_\chi = 30$ GeV and     
for $d = 0.05$. The same for other values of $d$ are almost identical. 
A number of features of the nature and
behaviour of the plots are apparent from Figs. 2 and 3 and they are the 
results of complicated dependence of rates on various factors, like 
variations of scalar cross sections with $m_\chi$ and $d$, the factor
$\xi$, the nuclear recoil energy etc. Let us make the following observations
in Figs 2 and 3. Firstly for a fixed value of $d$, say 0.05 (Fig 1a), the peak 
value of $\frac {\Delta R} {\Delta E}$ falls with increase of $m_\chi$. This
is mainly due to $1/(m_\chi m_{\rm red}^2)$ behaviour of rate equation 
and also due to decrease of $\sigma_{\rm scalar}$ with $m_\chi$. From 
Fig 1a, one can also see that the rate corresponding to lower $m_\chi$ 
falls off early. This feature has to do with the nature of variation 
of the expression $f_{\rm err}$ =        
$\l [ {\rm erf} \l ( \frac 
{v_{\rm min} + v_\oplus} {v_0} \r ) - {\rm erf} \l ( \frac
{v_{\rm min} - v_\oplus} {v_0} \r ) \r ]$ in Eq. (2.12) with 
recoil energy $E_R$. 
We have actually checked that the plot for $E_R$ vs $f_{\rm err}$ always 
remain lower for lower $m_\chi$ for a fixed value of $d$. It is also 
observed from Fig. 2a - 2b that the plots of 
$m_\chi = 30$, 50, 80, and 100 GeV's for $d=0.05$ (Fig. 2a) almost 
identical to those with $d=0.1$ (Fig. 2b) but the plots differs for 150 GeV
for these two values of d. 
This feature has its origin in the way $\xi$ is calculated. 
The relative decrease of $\sigma_{\rm scalar}$ for increase in $d$ for 
a particular $m_\chi$ is compensated by the relative increase in the 
value of $\xi$ (see Fig. 1 and discussion above regarding estimation of $\xi$).
For the case where this is not satisfied we find a difference in the nature
of variation of rate with energy. Thus  the plot corresponding to 
$m_\chi = 150$ GeV in Fig. 2b (where the value of $\xi$ is taken to be 1) 
is different from the corresponding plot in Fig. 2a. This therefore explains 
the deviations of a particular plot of a specific graph (a or b or c or d)
in Fig. 2 from the corresponding plot of other three graphs. 

From Fig. 2 we also observe that the defferential rate is very small for 
WIMP masses greater than 100 GeV while for lower WIMP mass it is not that 
small. As for example for $m_\chi= 50$ GeV, the differential rate 
at $E = 4.5$ keV is $\simeq 0.12$, whereas the same for a 150 GeV WIMP
varies from 0.082 for $d = 0.05$ to $3.3 \times 10^{-3}$ for $d = 0.4$. 
More so, for this $m_\chi$ (=50 GeV) we obtain a $d$ independent 
rate.  Also from Fig 1 and  above discussions, it can be said that 
for WIMP mass upto about 60 GeV, we may obtain $d$ independent rate. 

With the same values $\xi$, $m_\chi$ and $d$ as given in Table 1, we have 
also estimated the differential rate for a NaI detector. This detector
is a diatomic detector with detecting material consists of $^{23}_{11}$Na
and $^{127}_{53}I$ nuclei. In this case we have calculated 
first $\frac {dR} {dE_R}$ from Eq. (2.12) for $^{23}$Na and $^{127}$I 
separately and then use Eq. (2.15) for the computation of  
$\frac {\Delta R} {\Delta E}$ for NaI detector. The value of $v_0$ is kept
at 220 km/sec. The results are shown in four graphs of Fig. 4 (Fig. 4a - 
Fig. 4b) for four different values of $d$.  

In order to investigate the annual variation of WIMP detection rates 
we calculate the variation of total rate in one year. For this purpose
a representative value of degeneracy parameter $d = 0.2$ and WIMP mass
$m_\chi = 50$ GeV is considered. The ratio $\xi$ is read from Table 1.
In doing this, we vary $t$ in Eq. (2.8)
from 1 to 365 and for each value of $t$ (i.e. for each day from January 1)
and calculate the total rate by integrating the differential rate. In  
Fig. 5  we plot the results for the case of a $^{76}$Ge detector. 
Fig. 5 shows the sinusoidal behaviour of 
WIMP detection rate with respect to different days in a year. 
The prediction for maximum detection is in June and
the minimum is in December, as expected. We repeated the calculation
for NaI detector with similar results and they are shown in Fig. 6.

Although all the investigations above are for a fixed value of 
$v_0 = 220$ km/sec, we repeat the calculations for five representative
values of $v_0$ in the range 170 km/sec $\leq v_o \leq$ 270 km/sec. We calculate
the variation of differential rates with $E$ (in keV) 
for the five values $v_0$ = 170 km/sec, 200 km/sec,
230 km/sec 250 km/sec and 270 km/sec. with the value of $t$ in Eq. (2.8) 
fixed at $t_0$ (as earlier). The results for Ge detector and NaI detector 
are plotted in Fig. 7 and Fig. 8 respectively.  

\section {Conclusion}

We have considered lowest state Kaluza Klein particle (LKP) in universal 
extra dimension to be a candidate for cold dark matter. Unlike the 
supersymmetric particle, neutralino, which is a fermion, this Kaluza Klein 
particle is bosonic. We have predicted differential rates 
for detection of WIMP signals for  Ge and NaI detectors with KK particle
as Dark Matter candidate. The prediction is made by first estimating the
value of $\xi$ -- the ratio of local Dark Matter density to total 
local Dark Matter density and for different values of a degeneracy 
parameter $d$ (related to particle in universal extra dimension) and 
different values of WIMP mass. Due to diurnal motion of earth around the 
sun, the WIMP signals detected by earth bound detectors suffer an annual 
modulation with a maximum when the WIMP velocity with respect to the earth
is parallel to the earth's rotational velocity and a minimum when they are 
antiparallel. We have estimated the nature 
of annual modulation signal for WIMPs for the two types of detectors mentioned 
above. Lastly, we predicted the differential rate for five different
values of $v_0$ within its 90\% C.L. range to show the former's variation
with the circular velocity of the Local System. The new detectors like 
GENIUS, and running detectors like DAMA alongwith various other 
Dark Matter search programs of incresed sensitivity, can verify the 
possibility of LKP to be a candidate for Dark Matter.

\begin{center}
{\bf Figure Captions}
\end{center}

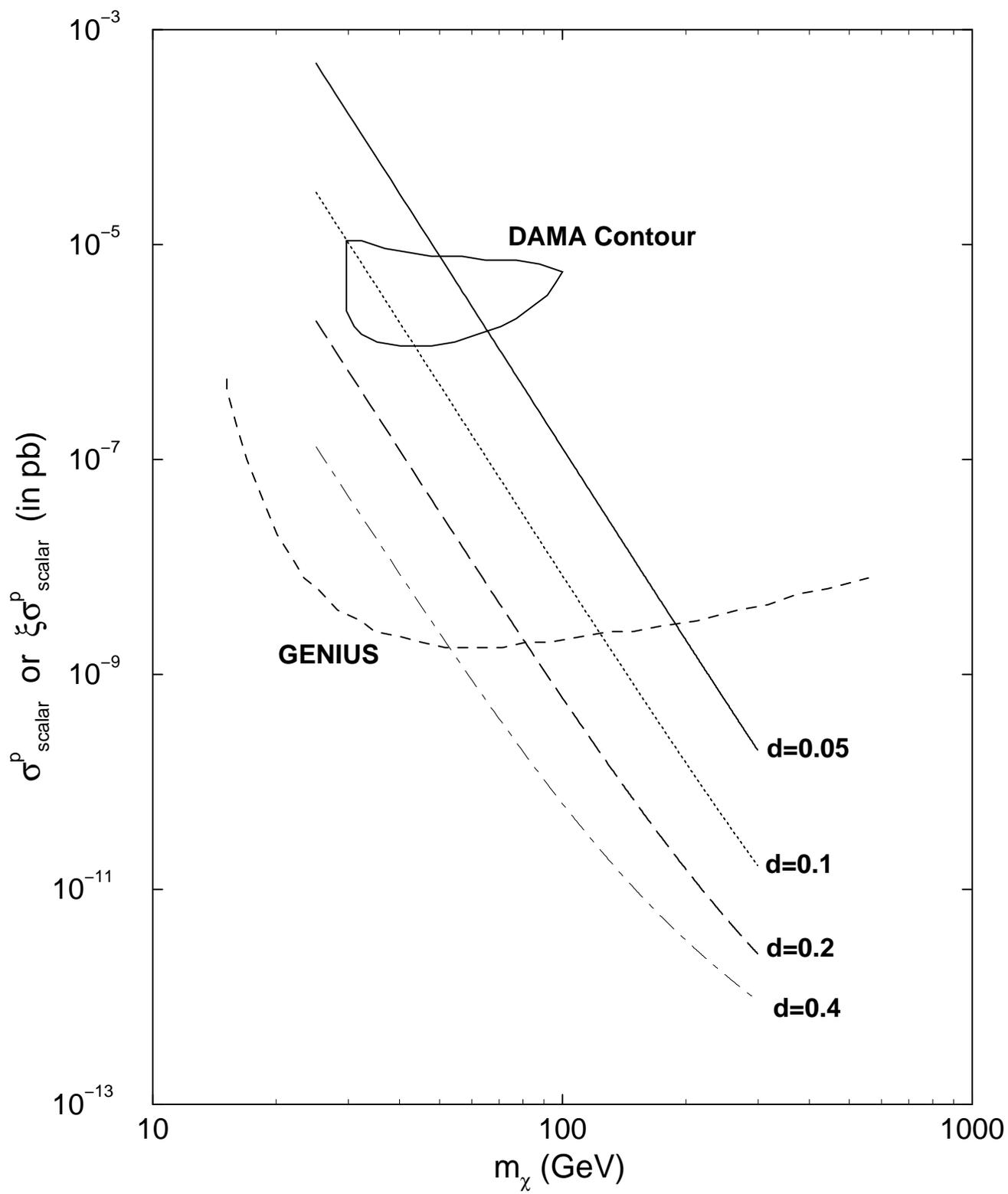
\noindent {\bf Fig. 1} Variation of $\sigma_{\rm scalar}^p$ with $m_\chi$
for LKP Dark Matter obtained from Eq. (2.13) for $d = 0.05,$ 0.1, 0.2 and 0.4
(see text). Also shown are 3$\sigma$
C.L. region given by DAMA/NaI direct detection results for WIMP 
and projected limit for GENIUS detector with 100 kg. Germenium.

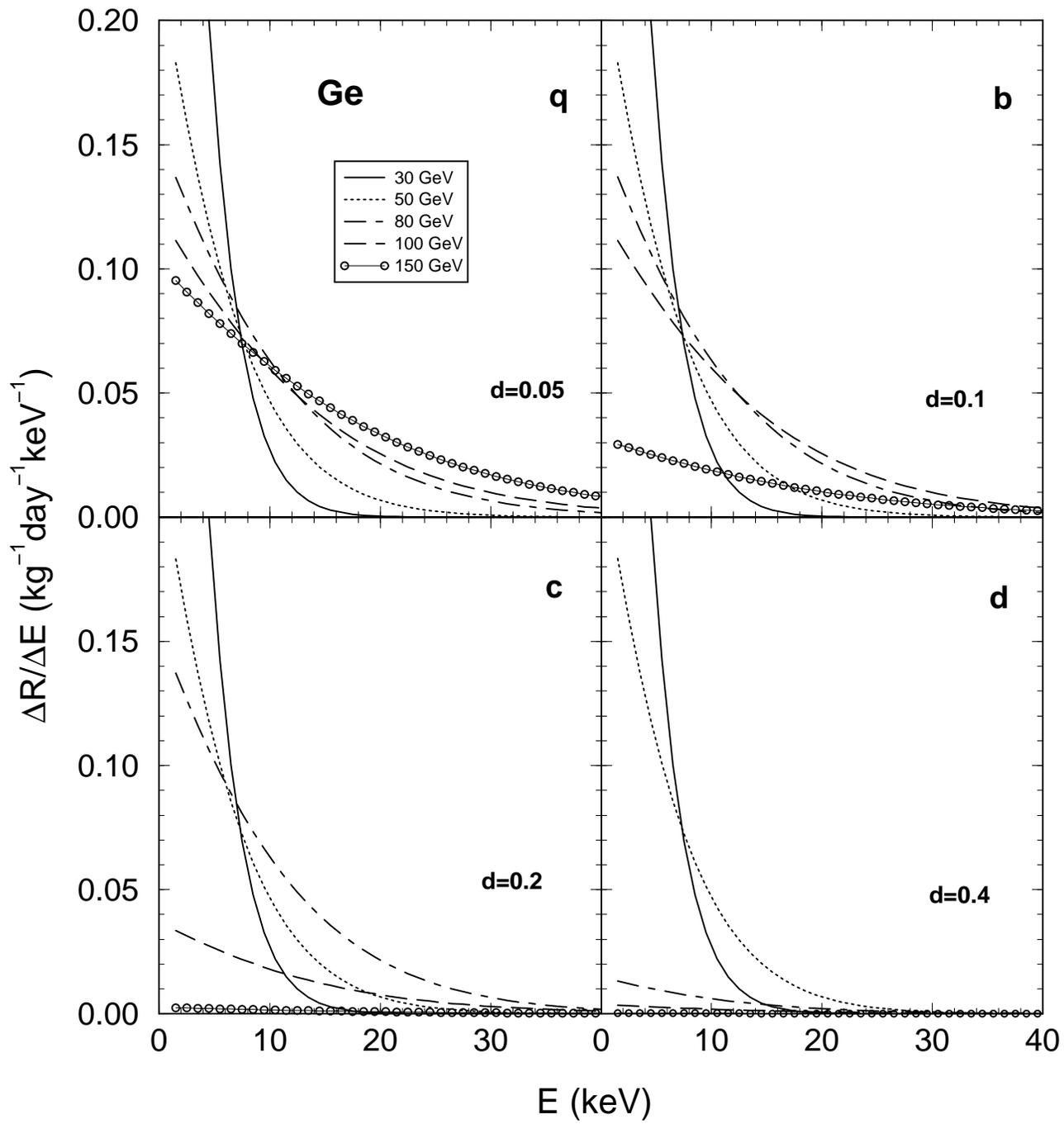
\noindent {\bf Fig. 2} Prediction of WIMP detection rates at $^{76}$Ge 
detector for five different WIMP masses
namely 30 GeV, 50 GeV, 80 GeV, 100 GeV and 150 GeV for different fixed
values of $d$. Figs. 2a, 2b, 2c and 2d correspond to $d=0.05$, 0.1, 0.2 
and 0.4 respectively. 

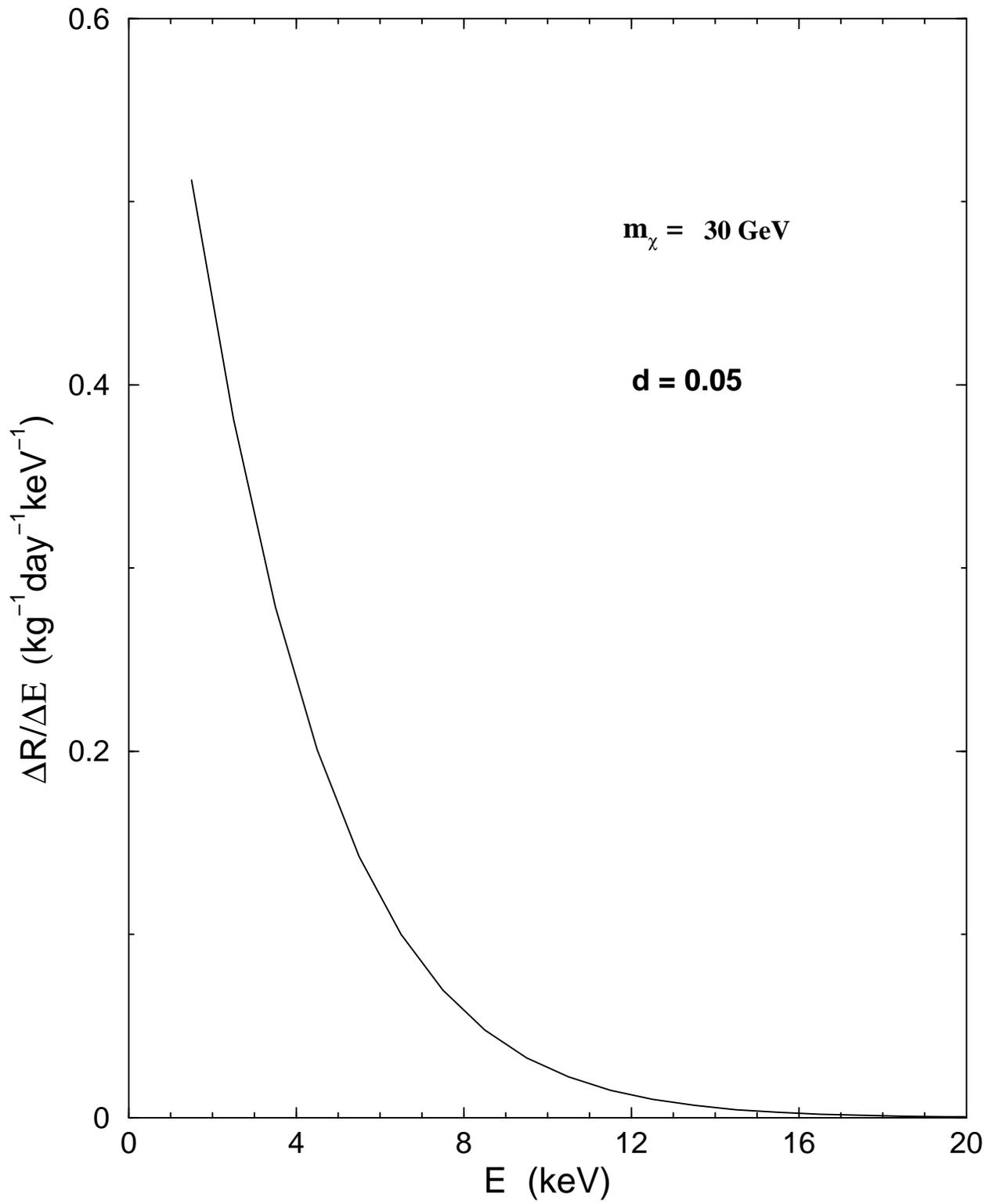
\noindent {\bf Fig. 3} WIMP detection rates for $^{76}$Ge with $d=0.05$
and $m_\chi = 30 GeV$ (see text). 

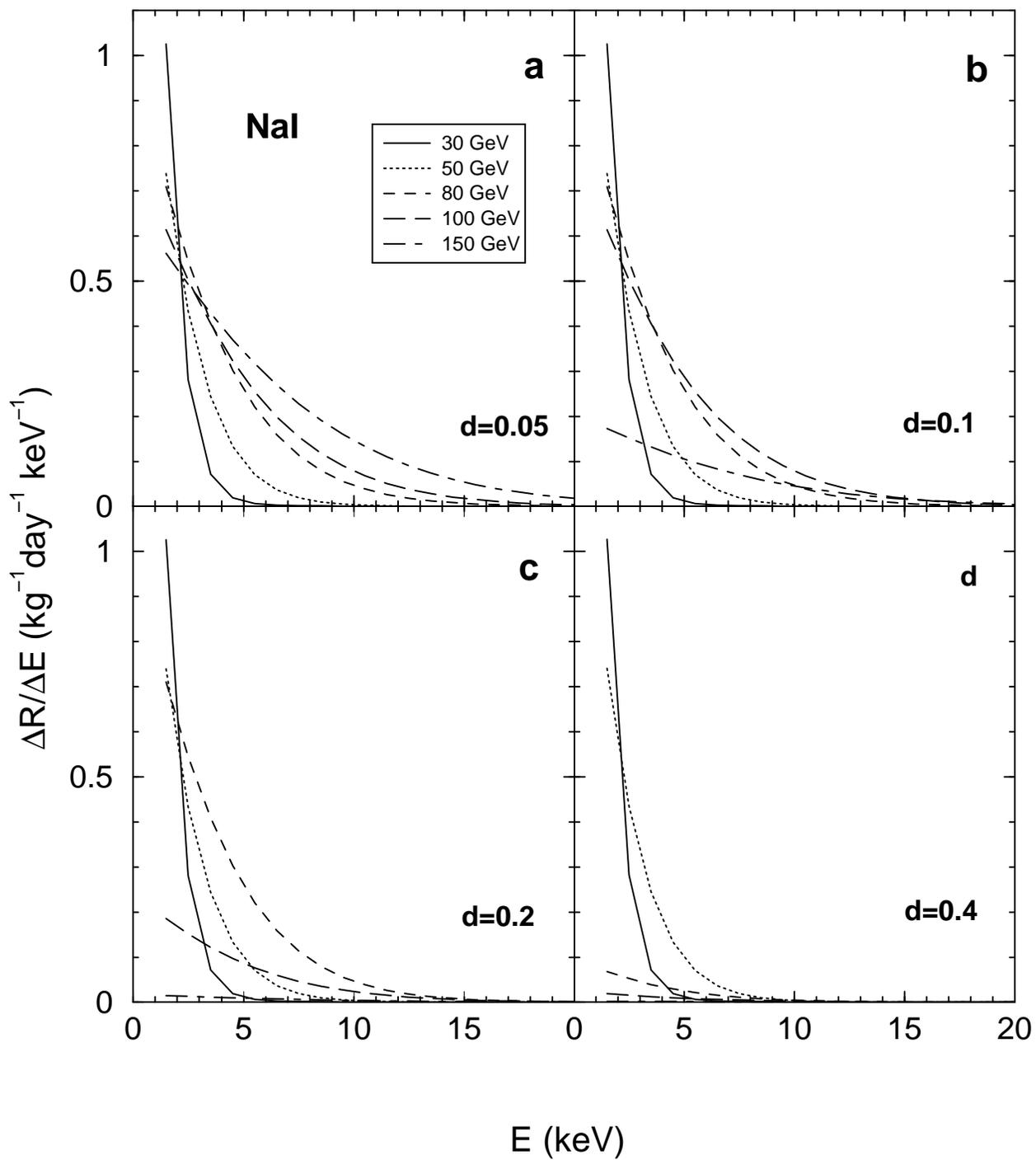
\noindent {\bf Fig. 4} Same as Fig. 2 but for NaI detector.

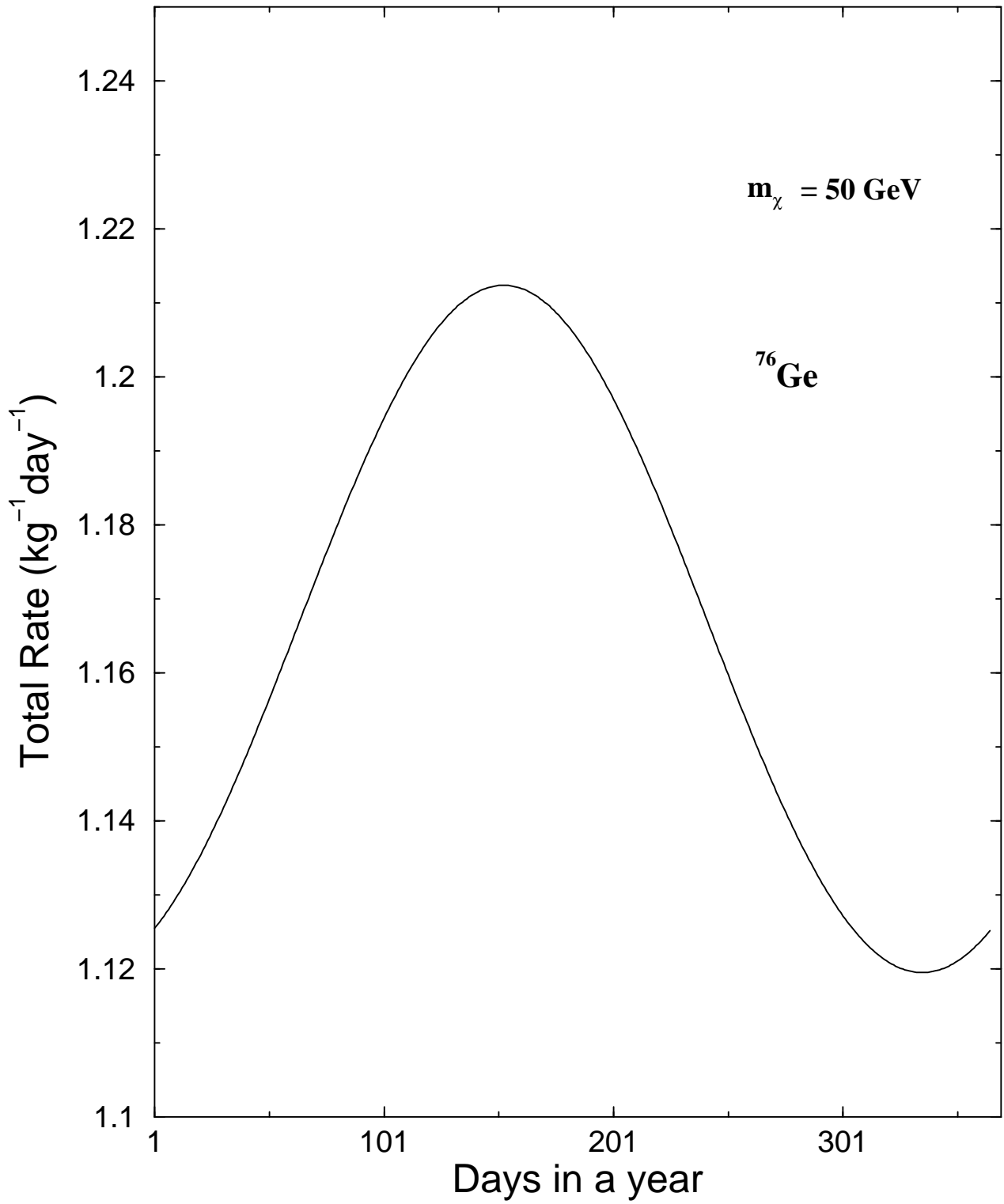
\noindent {\bf Fig. 5} Annual modulation of total WIMP detection 
rate per kg per day for $^{76}$Ge detector. 

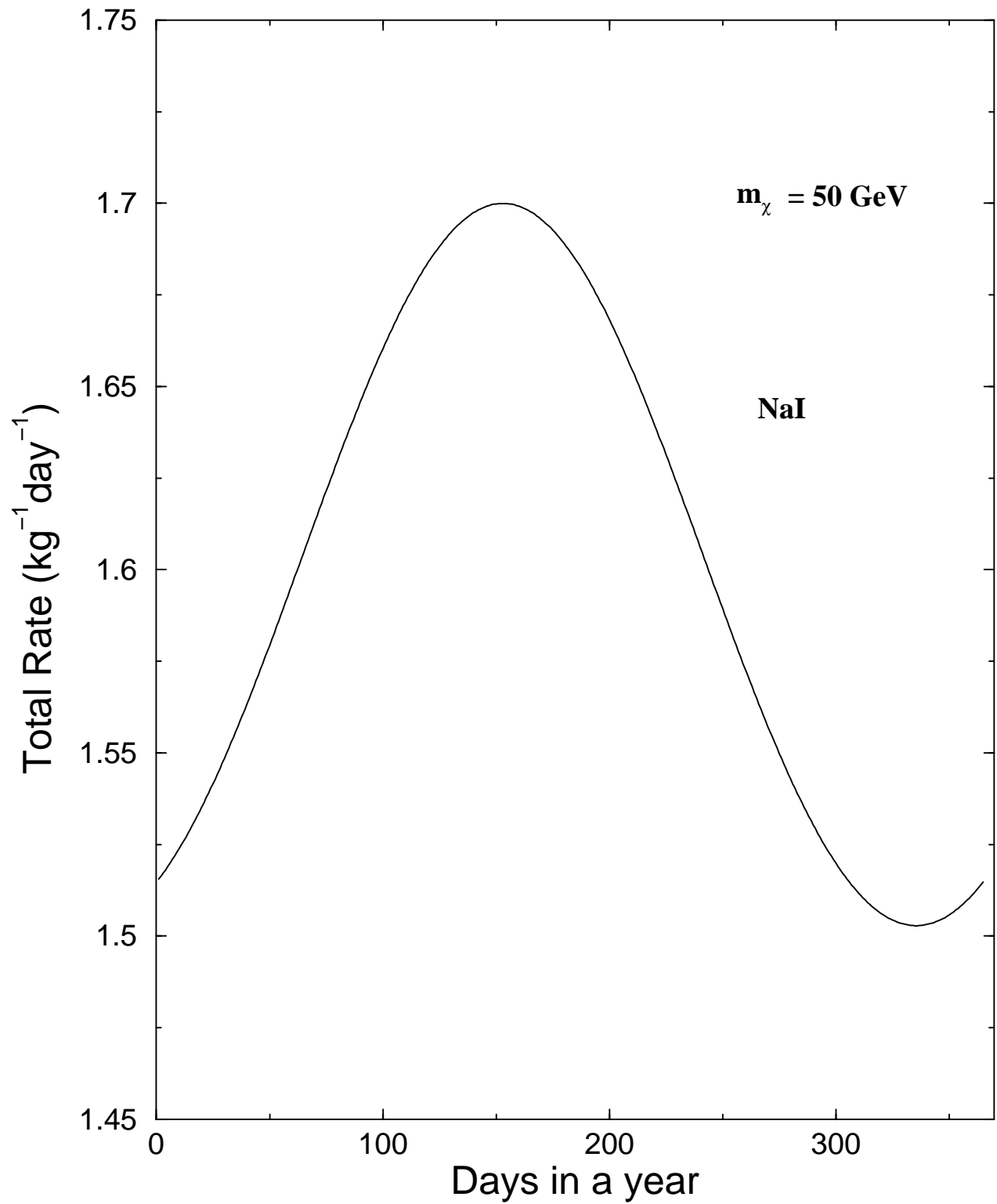
\noindent {\bf Fig. 6} Same as Fig. 5 but for NaI detector.

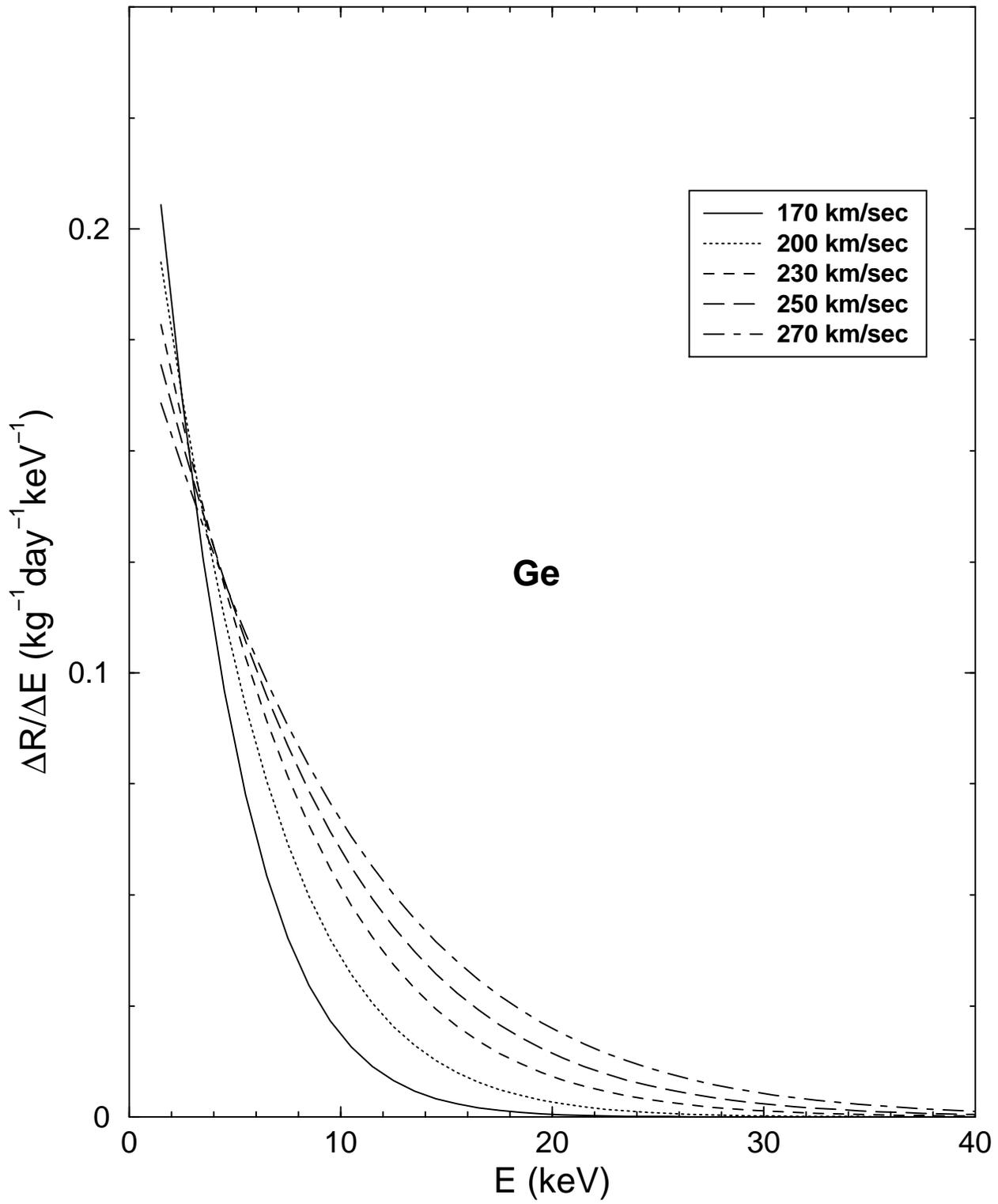
\noindent {\bf Fig. 7} WIMP detection rate prediction (per kg per day per 
keV) for different values of $v_0$ (see text) for $^{76}$Ge detector.

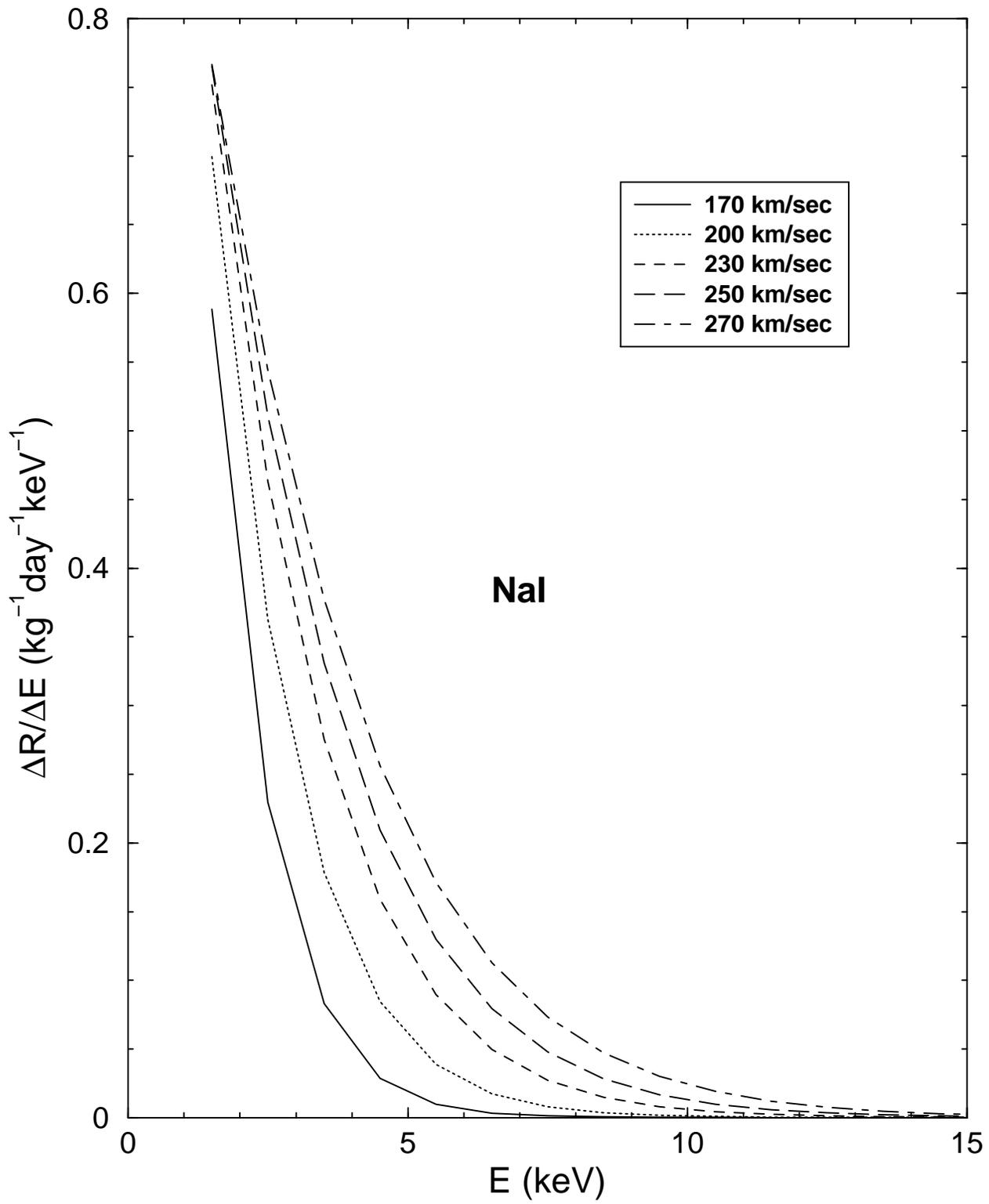
\noindent {\bf Fig. 8} Same as Fig. 7 but for NaI detector.   

\end{multicols}

\begin{thebibliography}{99}
\bibitem{bottino1} A. Bottino, F. Donato, N. Fornengo, S. Scolpel,
Phys. Rev. {\bf D59}, 095003 (1999).
\bibitem{bottino3} A. Bottino, V. de Alfaro, N. Fornengo, G. Mignola,
M. Pignone, Astropart. Phys. {\bf 2}, 67 (1994); 
A. Bottino, F. Donato, N. Fornengo, 
S. Scopel, Phys. Rev. {\bf D59}, 095004 (1999).
\bibitem{howard} H. Baer, M. Brhlik, Phys. Rev. {\bf D57}, 567 (1998).
\bibitem{edsejo} J. Edsejo, P. Gondolo, Phys. Rev. {\bf D56}, 1879 (1997).
\bibitem{utpal} U. Chattopadhyaya, A. Corsetti and P. Nath, Phys. Rev. 
{\bf D66}, 035003 (2002).
\bibitem{dm} D. Majumdar, J. Phys. G (2002), to appear.
\bibitem{cheng1} H.-C. Cheng, J.L. Feng, K.T. Matchev, eprint no. 
                 hep-ph/0207125.
\bibitem{cheng2} H.C. Cheng, eprint no. hep-ph/0206035.
\bibitem{kk} T. Kaluza, Sitzungsber. Preuss. Akad. Wiss. Berlin (Math. Phys.)
             {\bf K1}, 966 (1921); O. Klein, Z. Phys. {\bf 37}, 895 (1926)
             [Surveys High Energy Phys. {\bf 5}, 241 (1986)].
\bibitem{dama} R. Berbeiri et al, Phys. Lett. {\bf B533} 4 (2000). 
\bibitem{cdms} D. Abrams et al, CDMS collaboration eprint no. 
astro-ph/0203500.
\bibitem{edelweiss} A. Benoit et al, EDELWEISS collaboration, eprint no.
astro-ph/0206271.
\bibitem{klapdor0} H.V. Klapdor-Kleingrothaus et al, eprint no. hep-ph/0206151.
\bibitem{klapdor1} H.V. Klapdor-Kleingrothaus, B. Majorovits in 
{\it York 2000, The identification of dark matter} (2000), 
eprint no. hep-ph/0103079; 
H.V. Klapdor-Kleingrothaus, Nucl. Phys. Proc. Suppl. {\bf 110}, 364 (2002). 
and references therein.
\bibitem{bottino2} A. Bottino, V. de Alfaro, N. Fornengo, G. Mignola, 
S. Scolpel, Astropart. Phys. {\bf 2}, 77 (1994).
\bibitem{belli} P. Belli, R. Bernabei, A. Bottino, F. Donato, N. Fornengo, 
D. Prosperi, S. Scolpel, Phys. Rev. {\bf D61}, 023512 (2000).
\bibitem{brhlik} M. Brhlik, L. Roszkowski, Phys. Lett. {\bf B464}, 303 (1999).
\bibitem{jungman} G. Jungman, M. Kamionkowski, K Griest, Phys. Rep. {\bf 267},
                  195 (1996).
\bibitem{helm} R.H. Helm, Phys. Rev. {\bf D104}, 1466 (1956).
\bibitem{engel} J. Engel, Phys. Lett. {\bf B264}, 114 (1991).
\bibitem{leonard} P.J.T. Leonard, S. Tremaine, Ap. J. {\bf 353}, 486 (1990).
\bibitem{kochanek} C.S. Kochanek, Ap. J. {\bf 457}, 228 (1996).
\bibitem{isotope} {\it Table of Isotopes}, Eigth edition, Vol. 1, Ed.
R.B. Firestone, V.S. Shirley, John Wiley \& Sons, 1996.
\end{thebibliography}
\end{document}